# Quantum geometry induced anomalous chiral transport and hidden symmetry breaking in centrosymmetric 2*M*-$WS_2$

Hang Cui[1,#], Shao-Bo Liu[1,#,⁑], Erqing Wang[1,#], Mingxiang Pan[1], Yuqiang Fang[2,†], Ning Ma[1], Wenlong Liu[1], Di Chen[3], Yu Zhang[3], Yuanjun Song[3], Tingting Hao[3], Jiankun Li[3], Jian Cui[3], Ya Feng[3], Haiwen Liu[4,5], Fuqiang Huang[2], Huaqing Huang[1,‡], X.-C. Xie[1,6,7], Jian-Hao Chen[1,3,7,8,*]

[1] International Center for Quantum Materials, School of Physics, Peking University, Beijing, China

[2] School of Materials Science and Engineering, Shanghai Jiao Tong University, 800 Dongchuan Road, Shanghai, 200240, China

[3] Beijing Academy of Quantum Information Sciences, Beijing, China

[4] Center for Advanced Quantum Studies, department of Physics, Beijing Normal University, Beijing 100875, China

[5] Key laboratory of Multiscale Spin Physics, Ministry of education, Beijing Normal University, Beijing 100875, China.

[6] Institute for Nanoelectronic Devices and Quantum Computing, Fudan University; Shanghai, 200433, China.

[7] Hefei National Laboratory; Hefei, 230088, China

[8] Key Laboratory for the Physics and Chemistry of Nanodevices, Peking University, Beijing, China

Chirality, a widely existing material property in nature involving the breaking of the left-right symmetry, has profound influences in various fields of natural sciences. Nonlinear response, such as electronic magnetochiral anisotropy (eMChA), has been recognized as a sensitive probe for the effects of symmetry breaking and nontrivial quantum geometries in solids. So far, observations of eMChA have primarily been limited to inversion-symmetry broken materials. Here, we report a remarkable chiral transport in centrosymmetric candidate topological superconductor 2*M*-$WS_2$ flakes observed via second-harmonic generation under an out-of-plane magnetic field. More importantly, the eMChA becomes significant around the crossover temperature $T_{FL}$ ~ 25 K from the Fermi liquid (FL) to strange metal (SM) in the normal state, which interestingly echoes with the anomalously large Nernst response at the same temperature in bulk 2*M*-$WS_2$. These observations reveal a direct correspondence between the nonlinear response, Nernst response, and FL-SM transition in 2*M*-$WS_2$. Theoretical analysis indicates that nontrivial quantum geometry is behind the simultaneous response of eMChA and Nernst effects in 2*M*-$WS_2$ and the contribution from the orbital magnetic moment at the Fermi surface becomes significant during the FL-SM transition. Based on first-principles calculations, a thick-layer-sliding mechanism with minimal energy gain in $2M$-$WS_2$ provides one possibility for the generation of such nontrivial quantum geometry. The intertwined physics of remarkable eMChA, Nernst response, and FL-SM transition make $2M$-$WS_2$ a rare quantum platform to study the chiral transport and unexplored phenomena in strange metals, which may shed light on the trans-century, unresolved scientific issue in unconventional high-temperature superconductivity.

Two-dimensional transition metal dichalcogenides (TMDs) have garnered significant attention as fascinating quantum materials, demonstrating intricate interplay between charge, spin, orbital, and lattice degrees of freedom. These unique interactions may lead to intertwined physics of quantum states with similar energies, giving rise to exotic physical properties that holds great potential for next-generation quantum devices. In particular, centrosymmetric candidate topological superconductor $2M$-$WS_2$ has been attracting considerable interest for three main reasons: 1) distinctive crystal structure [1,2], 2) topologically nontrivial superconducting (SC) state [3-8], 3) exotic normal state [9-11]. Structurally, $2M$-$WS_2$ is a newly discovered metastable phase where the inversion symmetry is always preserved in both odd and even layers [1,4]. Bulk $2M$-$WS_2$ is formed with $1T'$-$WS_2$ single layers stacking along the *a*-axis direction with neighboring layers offset through a translation operation instead of a glide mirror operation as in the $1T'$-$WTe_2$ structure [1,4,5]. In its SC state, $2M$-$WS_2$ exhibits the highest SC transition temperature ($T_c$ = 8.8 K) among all TMDs [1], possibly related to strong and mode-selective electron-phonon coupling [1,10]. Besides, $2M$-$WS_2$ may be a rare intrinsic candidate topological superconductor, possessing anisotropic Majorana bound states [3], topological edge/surface states [5-8], and spin-orbit-parity coupled topological superconductivity [4], which could be the central component for fault-tolerant quantum computing. More importantly, in its normal state [9], $2M$-$WS_2$ exhibit an anomalously large Nernst effect right at the crossover between a Fermi-liquid (FL) to a strange-metal (SM) state around the quasiparticle coherence temperature $T_{FL}$ ~ 25 K. The anomalous Nernst peak at this crossover suggests a substantial change in carrier entropy when entering the strange-metal state with elusive Fermionic excitation, advancing the understanding of the mechanism of strange metal [9]. Recently, scanning tunneling microscopy (STM) has revealed novel paradigm-breaking charge orders in $2M$-$WS_2$ that decouple from lattice symmetry and adopt five distinct orientations across sample regions. Notably, their transition temperatures span 21-46 K with spatial variations and fall within the material's SM phase, indicative of correlation-driven or correlation-stabilizing phenomena[11]. However, the underlying essence of these observations has remained obscured due to the lack of additional experimental observations and theoretical understanding. In this work, we look into the symmetry aspects of $2M$-$WS_2$ to decode its intertwined physics or hidden orders around $T_{FL}$.

Symmetry plays a central role in both fundamental physics and practical applications by determining the properties of matter [12-14]. In this respect, nonlinear effect not only provides a sensitive probe into symmetry breaking and quantum geometries in solids [12,13,15,16] [Fig. 1], but also provides useful functionalities including rectification and frequency conversion [12,17-21]. Specifically, in conductive systems, when the electric field **E** is applied, the resulting current **j** can be written as: $\mathbf{j}=\sigma^{(1)}\mathbf{E}^1+\sigma^{(2)}\mathbf{E}^2+\sigma^{(3)}\mathbf{E}^3+\cdots$. The even-order terms vanish for centrosymmetric systems and only appear in non-centrosymmetric systems, directly reflecting the effects of symmetry breaking on electronic transport [12]. For example, in non-centrosymmetric conductor under magnetic field **B**, application of a low-frequency a.c. current **I** results in a resistance $R$ that depends on the sign of both **I** and **B** [Figs. 1(a) and 1(b)]. This phenomenon is known as nonreciprocal charge transport or eMChA [12-14,17-19,22-31], reflecting the simultaneous inversion/time-reversal symmetry breaking of the material. The resistance due to eMChA is typically expressed as: $R(\mathbf{B},\mathbf{I})=R_0+R^{1\omega}+R^{2\omega}=R_0(1+\mu^2\mathbf{B}^2+\gamma\mathbf{B}\cdot\mathbf{I})$ in chiral systems [12-14,26], or $R(\mathbf{B},\mathbf{I})=R_0+R^{1\omega}+R^{2\omega}=R_0[1+\mu^2\mathbf{B}^2+\gamma(\mathbf{P}\times\mathbf{B})\cdot\mathbf{I}]$ in polar systems [16,17,32-35]. Here $R_0$ represents the resistance at zero magnetic field, $R^{1\omega}$ denotes the normal magnetoresistance, $R^{2\omega}$ corresponds to the eMChA depending on both **B** and **I**, $\gamma$ is the eMChA coefficient, and **P** is the axis of the characteristic "polar" unit vector. Due to perturbative nature, the **B**-induced first-harmonic/second-harmonic voltage ($V^{1\omega}/V^{2\omega}$) depends linearly/quadratically on **I**, respectively [Fig. 1(c)]. So far, eMChA have mainly been observed in inversion-symmetry broken systems, such as polar semiconductor/metal/superconductor systems [16,17,32-35] [Fig. 1(d)], trigonal TMDs systems [12,36] [Fig. 1(e)], and chiral systems [12-14,23,26,31] [Fig. 1(f)]. Recently, a report of eMChA in centrosymmetric kagome-metal $CsV_3Sb_5$ [14,37] suggests additional symmetry breaking, potentially arising from chiral orbital loop currents [14] or nontrivial quantum geometry [38] [Fig. 1(g)]. Besides, the eMChA has also been observed in centrosymmetric helimagnet $\alpha$-$EuP_3$ and antiferromagnet $PdCrO_2$, due to the inversion-symmetry breaking caused by the chiral spin texture/structure [39,40]. Therefore, searching for eMChA in centrosymmetric systems is a challenging but highly intriguing direction, which could help to reveal subtle inversion-symmetry breaking or hidden orders of matter.

In this study, we provide the first report on a remarkable eMChA in the centrosymmetric candidate topological-superconductor 2*M*-$WS_2$ flakes. More interestingly, the significant nonlinear response, anomalously large Nernst response, and FL-SM transition simultaneously occur around a crossover temperature $T_{FL}$ ~ 25 K in its normal state, pointing to a direct correspondence between them. Theoretical analysis indicates that nontrivial quantum geometry is behind the simultaneous response of eMChA and Nernst effects in 2*M*-$WS_2$ due to possible temperature-induced reconstruction of the Fermi surface [41]. Based on first-principles calculations, a thick-layer-sliding mechanism with minimal energy gain in 2*M*-$WS_2$ provides a likely route for the generation of such nontrivial quantum geometry. Overall, the intertwined physics (i.e., remarkable eMChA, Nernst response, and FL-SM transition) make 2*M*-$WS_2$ a rare quantum platform to study the chiral transport and unexplored phenomena in strange metal, which may provide new perspective on the trans-century, unresolved scientific issue of strange metal in high-temperature superconductors [42-46], heavy fermion systems [47], and other flat-band materials[48-50].

High-quality 2*M*-$WS_2$ single crystals were synthesized using the topochemical method of $K^+$ deintercalation from $K_{0.7}WS_2$ crystals [1] (see Supplementary Material, Note 1 for details). The obtained 2*M*-$WS_2$ crystals grew mainly along the *c*-axis direction to form a natural thin narrow ribbon [Figs. 2(a) and 2(b)] [9]. To obtain high-quality electrical transport data, we fabricated thin-flake 2*M*-$WS_2$ Hall bar devices [Fig. 2c] using top-down mechanical cleavage onto the $SiO_2$ (285nm)/Si substrate, electron beam lithography and electron beam evaporation techniques (details in Supplementary Material, Note 1 and Fig. S1). To measure eMChA of 2*M*-$WS_2$ devices (five devices with thickness from 35 to 150 nm have been measured, see Supplementary Fig. S2), we bias the devices with a harmonic current at a fixed low frequency (17.777 Hz) along the *c*-axis direction and record both the first-harmonic and second-harmonic signals with lock-in techniques (see Supplementary Material, Note 1 for details).

Figures 2(d)-2(i) and Supplementary Figs. S3&S4 are the eMChA and first-harmonic results of 2*M*-$WS_2$ flake (device #1 with a thickness of 130 nm). The insets show the electrode geometry and measurement conditions. Electrical current is injected from the source (S) to the drain (D) electrode and the voltage is measured between the *V*+ and *V*- electrodes. Figures 2(d) and 2(e) show the magnetic field-dependent longitudinal first-harmonic voltage $V_{xx}^{1\omega}$ and second-harmonic voltage $V_{xx}^{2\omega}$ measured under $T$ = 20 K and $I_{ac}$ = 500 $\mu$A, respectively. Figures 2(f) and 2(g) show the magnetic

field-dependent $V_{xx}^{1\omega}$ and $V_{xx}^{2\omega}$ measured under $T$ = 20 K and $I_{ac}$ = 500 $\mu$A when switching the measurement electrodes $V$+ and $V$-. Figures 2(h) and 2(i) show the $V_{xx}^{1\omega}$ and $V_{xx}^{2\omega}$ versus a.c. current amplitude measured under **B** = 9 T and $T$ = 10, 20, 30, 40, and 80 K, respectively. From a phenomenological point of view, the current-voltage relationship of the non-reciprocal transport can be written as: $V(\mathbf{B,I})=V_0+V^{1\omega}+V^{2\omega}=R_0I_0+R_0I_0\mu^2\mathbf{B}^2+\gamma R_0I_0^2\mathbf{B}\cdot\hat{\mathbf{I}}$ (Eqn.1) or $V(\mathbf{B,I})=V_0+V^{1\omega}+V^{2\omega}=R_0I_0+R_0I_0\mu^2\mathbf{B}^2+\gamma R_0I_0^2(\mathbf{P}\times\mathbf{B})\cdot\hat{\mathbf{I}}$ (Eqn.2), where $V$ is voltage, $I_0$ is current, $R_0$ is the resistance that does not change with the current, **P, B**, **I** and $\gamma$ are the same parameters as previously defined. Indeed, the experimentally measured $V_{xx}^{1\omega}$ is symmetric about **B** [Fig. 2(d)], while $V_{xx}^{2\omega}$ is antisymmetric about **B** [Fig. 2(e)]; $V_{xx}^{1\omega}$ is not dependent on $\hat{\mathbf{I}}$ [Fig. 2(f)], while $V_{xx}^{2\omega}$ is antisymmetric about $\hat{\mathbf{I}}$ [Fig. 2(g)]; $V_{xx}^{1\omega}$ depends quadratically on **B** [Fig. 2(d) and Fig. 2(f)], while $V_{xx}^{2\omega}$ depends linearly on **B** [Fig. 2(e) and Fig. 2(g)]; $V_{xx}^{1\omega}$ depends linearly on $I_0$ [Fig. 2(h)], while $V_{xx}^{2\omega}$ depends quadratically on $I_0$ [Fig. 2(i)]. Thus, strong and well-defined non-reciprocal transport is found in the putative centrosymmetric 2$M$-$WS_2$, revealing a hidden symmetry breaking state in the material.

Crucial information about such hidden symmetry breaking state in 2$M$-$WS_2$ is revealed by the dependence of the eMChA under various parameters. In particular, the magnetic field angular dependent $V_{xx}^{2\omega}$ in the $xz$-, $yz$- and $xy$-plane under $I_{ac}$ = 500 $\mu$A and $T$ = 20 K are shown in Figs. 3(a)-3(c), respectively. In both the $xz$- and $yz$-planes, $V_{xx}^{2\omega}$ shows a $\cos\theta$ or $\cos\alpha$ dependence ($\theta$ is measured from the $z$ axis in the $xz$-plane, and $\alpha$ is measured from the $z$ axis in the $yz$-plane), while $V_{xx}^{2\omega}$ remains essentially zero in the $xy$-plane. Therefore, it is surprising that the second-harmonic signal in the centrosymmetric 2$M$-$WS_2$ flake obey Eqn. (2) stated in the previous section [29,33,34,51]. Since **I** is along the $x$ direction ($c$ axis), this result [Figs. 3(a)-3(c)] indicates that **P** is along the $y$ direction ($b$ axis). The characteristic behavior observed in the centrosymmetric 2$M$-$WS_2$ flake is thus similar to that reported in polar systems such as the gated oxide interface $LaAlO_3/SrTiO_3$, ferromagnetic polar metal $Ca_3Co_3O_8$ and topological semimetal $ZrTe_5$ thick flake, which exhibit spontaneous inversion symmetry breaking [29,33,34,51]. Figure 3(d) shows temperature dependent $V_{xx}^{2\omega}$ in the normal state measured at $I_{ac}$ = 500 $\mu$A in vertical **B** = 9, 7, 5, 3, -3, -5, -7 and -9 T. With temperature cooling from 80 K to 10 K, the $V_{xx}^{2\omega}(T)$ curves are antisymmetric on **B** and also display a peak feature around $T \sim$ 25 K. Figure 3(e) shows magnetic-field dependent $V_{xx}^{2\omega}$ measured at $I_{ac}$ = 500 $\mu$A and $T$ = 20, 30, 40, 70, 80 and 100 K, whose results are consistent with Fig. 3(d). To evaluate the strength of eMChA in 2$M$-$WS_2$, we measured the temperature-dependent eMChA coefficient $\gamma$, which is extracted from Supplementary Figs. S4(a) and S4(b) using the formula $\gamma=\frac{2R^{2\omega}}{R^{1\omega}|B||I|}$. As shown in Fig. 3(f), in the normal state $2M$-$WS_2$, the eMChA coefficient $\gamma$ is up to 0.5 $T^{-1}A^{-1}$, which is much higher than those observed in typical polar conductors, such as Bi helix (~$10^{-3}$ $T^{-1}A^{-1}$) [35], chiral organic conductor (~$10^{-2}$ $T^{-1}A^{-1}$) [23], and 2DEG in Si FET (~$10^{-1}$ $T^{-1}A^{-1}$) [52]. The above eMChA results observed in $2M$-$WS_2$ can be repeated in five different devices with thickness varying from ~ 35 nm to 150 nm (see Supplementary Figs. S5-S9). Extrinsic origins of the measured eMChA are ruled out, such as different driving frequency (device #4, see Supplementary Fig. S10(a)), Joule heating (device #3, see Supplementary Fig. S10(b)), and uneven voltage distribution caused by non-Ohmic contacts (see Supplementary Figs. S5(d) and S7(d)), leaving the only possibility to be intrinsic eMChA in $2M$-$WS_2$.

In order to better understand the origin of eMChA and its association with the intertwined physics in $2M$-$WS_2$, we summarized the data of our first/second-harmonic generation and other exotic physical phenomena occurring near 25 K in the material [9] in Fig. 4. The phase diagram of $2M$-$WS_2$ in the normal state [Fig. 4(a)] can be divided into three regions: 1) Fermi-liquid (FL) region highlighted by pink color (8.8 K < $T$ < 25 K); 2) Strange-metal (SM) region highlighted by light green color (25 K < $T$ < 120 K); 3) Bad-metal region highlighted by gray color (180 K < $T$ < 300 K). There are four different types of data in the phase diagram: 1) The blue circles represent the temperature dependent scattering rate extracted from resistivity and Hall resistivity, denoted as $\Gamma_{tr}(m^*/m_e)$. Three blue lines across the blue circles are quadratic fitting (8.8 K < $T$ < 25 K, FL behavior), linear fitting (25 K < $T$ < 120 K, SM behavior), and linear fitting (180 K < $T$ < 300 K, bad metal behavior), respectively [9]. 2) The green diamonds with green line (fitting to ~$1/T$) represent the temperature-dependent Hall coefficient $R_H$, consistent with the strange metal behavior. 3) The two solid red lines represent temperature dependent second-harmonic voltage $\Delta V_{xx}^{2\omega}=\Delta V_{xx}^{2\omega}(9\text{ T})-\Delta V_{xx}^{2\omega}(-9\text{ T})$ measured under $I_{ac}$ = 500 $\mu$A (the upper red line, $T$ = 8.8 K to 80 K) and 200 $\mu$A (the lower red line, $T$ = 8.8 K to 300 K), respectively; 4) The black squares with black line represent the temperature dependent Nernst coefficient $\nu$, which is extracted from the literature [9]. The source data for 1) to 3) is plotted in Supplementary Fig. S4(a)-S4(c), as well as in Supplementary Figs. S11 and S12.

The above phase diagram conveys a lot of information, with three main points: 1) Possible hidden symmetry breaking is revealed by the unexpected appearance of substantial eMChA in the putative centrosymmetric $2M$-$WS_2$ [red lines in Fig. 4(a)]; 2) Common underlying physical origin of the eMChA and the anomalous Nernst response is inferred as both signals peak around the FL-SM transition temperature $T_{FL}$ ~ 25 K [Fig. 4(a)] with the same magnetic field/temperature dependence [Figs. 4(b) and 4(c)]; 3) A rare nonmonotonic hump behavior of the eMChA at around $T_{FL}$, where the charge carriers transform from well-defined quasiparticles in the FL state to fermionic excitation in the strange metal state. Previously, only monotonically increasing second-harmonic signal with lowering temperature is reported, such as topological kagome metal $CsV_3Sb_5$ [14], topological semimetal $ZrTe_5$ [29] and polar Weyl semimetal $WTe_2$ [16]. Moreover, this is the first report of the nonlinear electrical response in strange metal, which may advance the understanding of the mechanism of this quantum state.

Now we discuss the possible origin of eMChA in centrosymmetric 2*M*-$WS_2$. The emergence of eMChA usually points to a lack of inversion symmetry [12,13,15,16], raising the question on the specific inversion symmetry breaking mode in 2*M*-$WS_2$. The striking coincidence of the eMChA and the anomalously large Nernst response in their temperature and magnetic-filed dependence [Figs. 4(b) and 4(c)] strongly indicates a common underlying mechanism, which we shall argue that significant quantum geometry at the Fermi surface is involved. From a physical standpoint, orbital magnetization in a topological material arises not only from localized magnetic moments, but also from boundary currents that circulate at the sample's edges. These edge currents are governed by the imaginary part of the quantum geometry at the Fermi surface and can produce a significant magnetic moment, thereby giving rise to a pronounced anomalous Nernst response [53] as described by Eq. (11) in Supplementary Note 2 for details. At the same time, in systems where inversion symmetry is broken, the quantum geometry at the Fermi surface can also drive the eMChA effect. This effect, too, can be understood as a consequence of chiral edge currents. Therefore, both the anomalous Nernst and eMChA effects are expected to be enhanced in regimes where the quantum geometry near the Fermi level is particularly large.

Utilizing first-principles calculations within the framework of density functional theory, we systematically examine possible mechanisms responsible for inversion-symmetry breaking and generation of non-trivial quantum geometry in 2*M*-$WS_2$, which can be seen from Supplementary Note 3 for details. First, we exclude the possibility of achieving inversion-symmetry breaking through a phase transition from 2*M*-$WS_2$ to $T_d$-$WS_2$ due to the substantial activation energy barrier (2 eV to 3.5 eV) impossible to achieve in our experimental conditions (details in Supplementary Fig. S14). Second, we exclude the possibility of breaking the inversion symmetry of 2*M*-$WS_2$ from *C*2/*m* to *C*2 or *C*m phases, because the two additional displacement modes involved in the *C*2 or *C*m phases are energetically unfavorable (details in Supplementary Figs. S15 and S16). Finally, based on first-principles calculations, we proposed a possible thick-layer-sliding mechanism with multiple possible sliding configurations and arbitrarily small sliding. Such mechanism involves very small activation energy to slide thick layers in 2*M*-$WS_2$, which subtly perturbs the local symmetry while preserving the parent 2*M*-$WS_2$ polytype framework. This mechanism initiates non-trivial quantum geometry in the material with an effective inversion symmetry breaking, with its characteristic conductivity tensor primarily governed by the orbital Zeeman effect and non-trivial quantum geometry (details in Supplementary Figs. S17-S20). The magnitude of eMChA, calculated based on this mechanism, is found to be comparable to our experimental results (details in Supplementary Note 4.1, Supplementary Figs. S21-S25 and Supplementary Table 2). The summarized device-dependent eMChA and FL-SM crossover (Supplementary Table 2) and discussions (details in Supplementary Note 4) seemingly further support the close link between them.

The next naturally arise question is: why does the non-trivial quantum geometry at the Fermi surface become especially significant around 25 K? One plausible explanation lies in a temperature-induced reconstruction of the Fermi surface [41], i.e., a hidden order with electronic-symmetry breaking, which could lead to the formation of quantum geometry "hot spots" near the Fermi energy, as confirmed by our calculations of the orbital magnetic moment contributions (Supplementary Fig. S27 and Supplementary Note 4.2). These localized regions of enhanced orbital magnetic moment would, in turn, amplify both the eMChA and anomalous Nernst responses observed experimentally.

In summary, we report a remarkable eMChA in the centrosymmetric candidate topological-superconductor 2*M*-$WS_2$ flake. More interestingly, the significant nonlinear effect, anomalously large Nernst response, and FL-SM transition simultaneously occur around a crossover temperature $T_{FL} \sim 25$ K in its normal state, pointing to a direct correspondence between them. Theoretical analysis indicates that nontrivial quantum geometry is behind the simultaneous response of eMChA and Nernst effects in 2*M*-$WS_2$. Based on first-principles calculations, a thick-layer-sliding mechanism with minimal energy gain in 2*M*-$WS_2$ provides one possibility for the generation of such nontrivial quantum geometry. Overall, the intertwined physics (i.e., remarkable eMChA, large Nernst response, and FL-SM transition) make 2*M*-$WS_2$ a rare quantum platform to study the chiral transport and unexplored phenomena in strange metal, which may provide new insight into the trans-century, unresolved scientific issue of strange metal in high-temperature superconductor [42-46], heavy fermion systems [47], and other flat-band materials [48-50].

Acknowledgments.—This project has been supported by the National Key R&D Program of China (Grant No. 2024YFA1409001), the Innovation Program for Quantum Science and Technology (Grant No. 2021ZD0302403), the National Natural Science Foundation of China (Grant Nos. 92265106, 12304537). Shanghai Rising-Star Program (Grant Nos. 23QA1410700). The National Key R&D Program of China (Grant No. 2021YFA1401600), the National Natural Science Foundation of China (Grants No. 12474056), the Science and Technology Commission of Shanghai Municipality (Grant No.24LZ1401000)， and the Training Program of the Major Research Plan of the National Natural Science Foundation of China (92565109). J.-H.C. acknowledges technical supports form Peking Nanofab.

H.C., S.-B. L., and E.Q. W. contributed equally to this work.

*Corresponding Authors:

Jian-Hao Chen (chenjianhao@pku.edu.cn),

⁑Corresponding Authors:

Shao-Bo Liu (liushaobopku@pku.edu.cn)

†Corresponding Authors:

Yuqiang Fang (fangyuqiang@mail.sic.ac.cn)

‡Corresponding Authors:

Huaqing Huang (huaqing.huang@pku.edu.cn )

## Figures

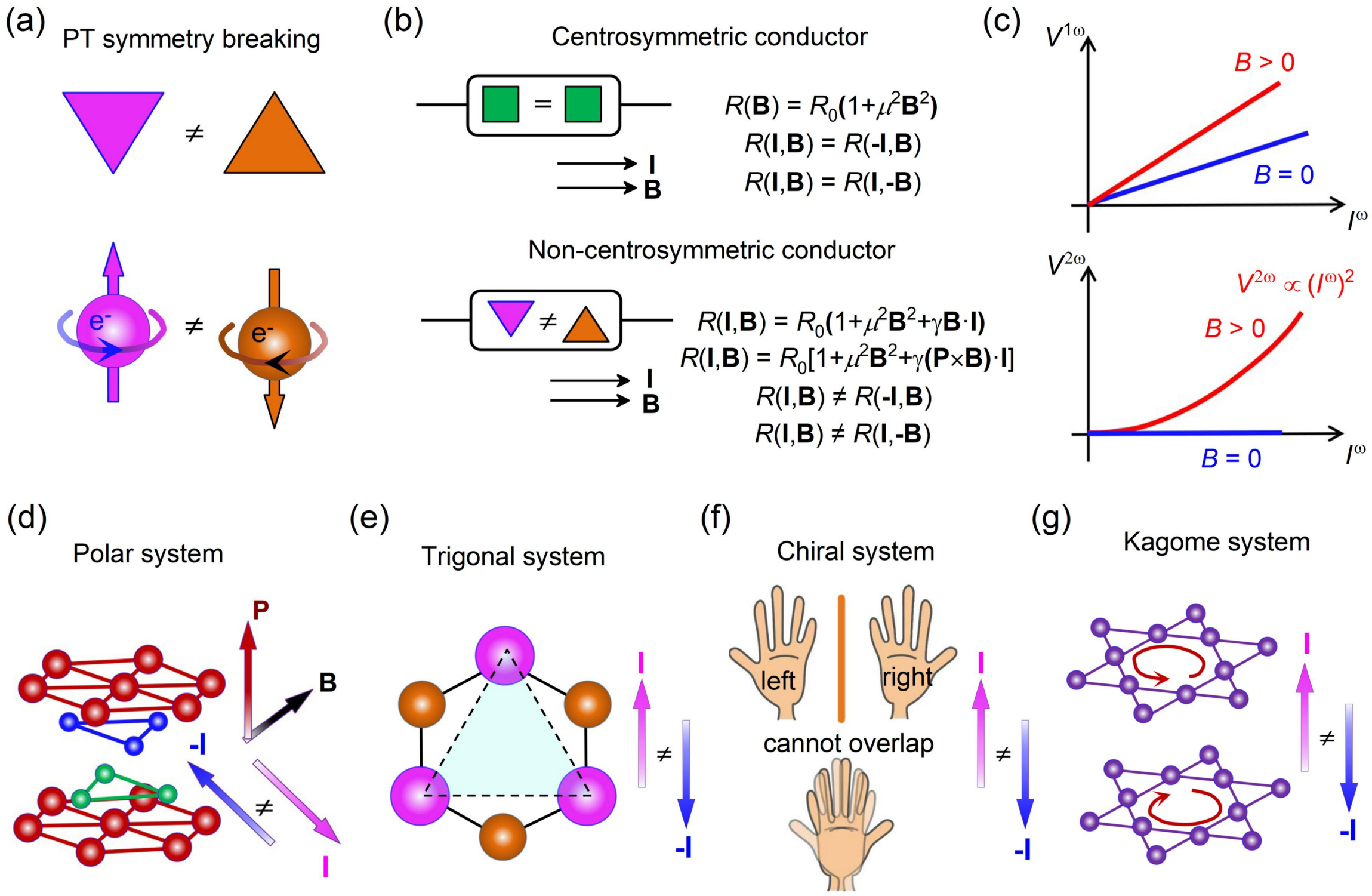

FIG. 1. Nonreciprocal transport under magnetic field in non-centrosymmetric crystals. (a) Spatial inversion and time reversal symmetry breaking in solids. (b) Illustration of electrical resistance of centrosymmetric and non-centrosymmetric conductors within the low-frequency a.c. current. (c) $I(V)$ curve for a non-centrosymmetric conductor. In the a.c. measurement, the field-induced first-harmonic voltage (up), $V^{1\omega}$, depends linearly on the a.c. current $I^{\omega}$; and the field-induced second-harmonic voltage (up), $V^{2\omega}$, depends quadratically on the a.c. current $I^{\omega}$. (d)-(f) Typical non-centrosymmetric systems, including the polar system in (d), monolayer trigonal system in (e), and chiral system in (f). (g) The structurally centrosymmetric kagome system ($CsV_3Sb_5$) with a spontaneously extra symmetry breaking at low temperatures due to the chiral orbital loop current. In each system, nonreciprocal transport satisfies the characteristic selection rule reflecting the symmetry breaking.

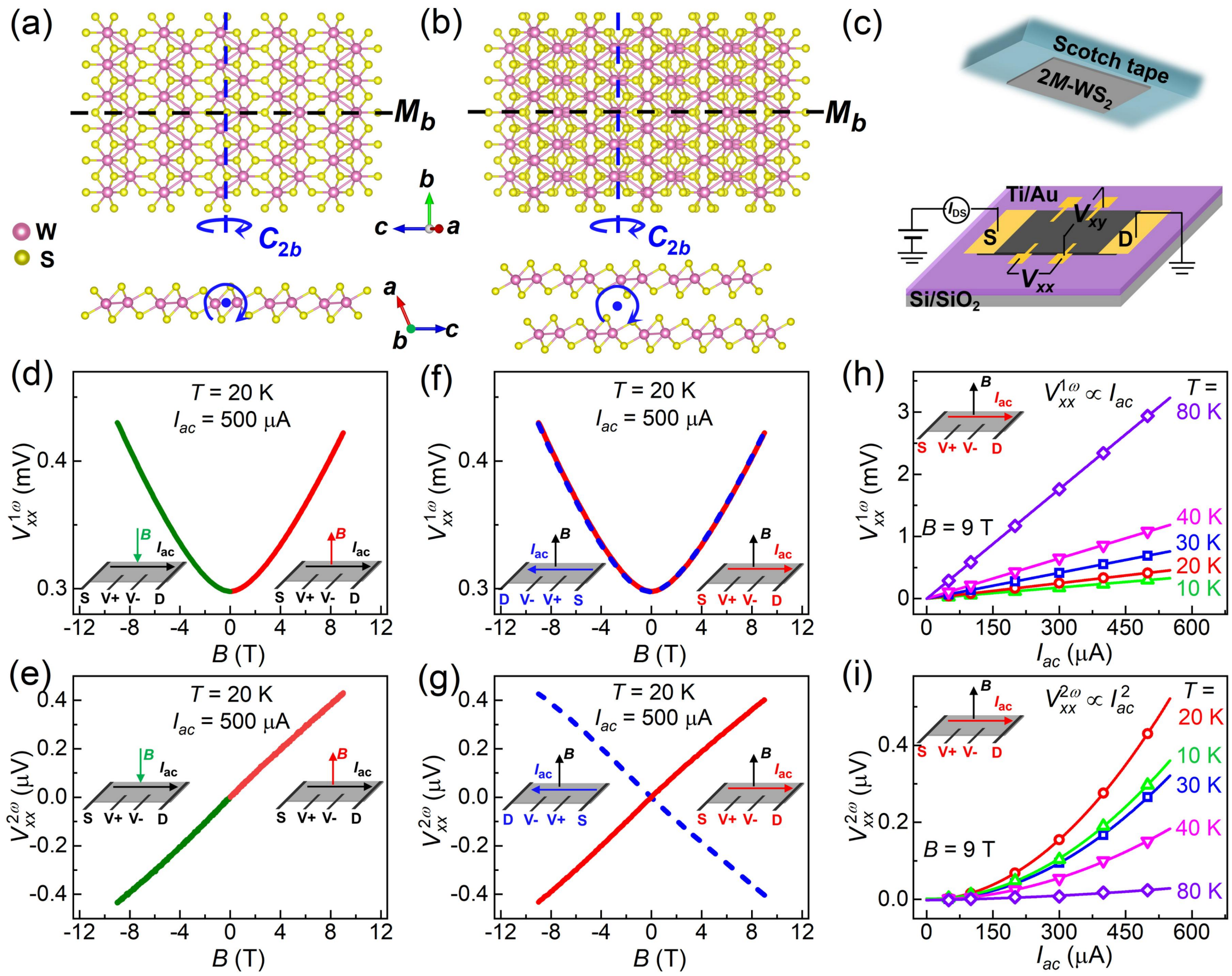

FIG. 2. Electrical magnetochiral anisotropy (eMChA) in centrosymmetric 2*M*-$WS_2$ flake. (a)-(b) The symmetries in monolayer and bilayer 2*M*-$WS_2$, exhibiting a mirror plane $M_b$ and a screw rotation symmetry $C_{2b}$, respectively. As a result, there always exists a global inversion centre both in odd- and even-layer 2*M*-$WS_2$. (c) Schematic diagram of cleavage, processing and measurement of few-layer 2*M*-$WS_2$ micro-nano devices. (d)-(e) Magnetic field-dependent first-harmonic voltage $V_{xx}^{1\omega}$ and second-harmonic voltage $V_{xx}^{2\omega}$ measured under $T$ = 20 K and $I_{ac}$ = 500 $\mu$A, respectively. The insets show the electrode geometry for the negative (green) and positive (red) magnetic field directions. The current is injected from the source (S) to the drain (D) electrode and the voltage is measured between the $V$+ and $V$- electrodes. (f)-(g) Magnetic field-dependent $V_{xx}^{1\omega}$ and $V_{xx}^{2\omega}$ measured under $T$ = 20 K and $I_{ac}$ = 500 $\mu$A when changing the current directions, respectively. The insets show the electrode geometry for the forward current (red) and backward current (blue) directions. The current is injected from the source (S) to the drain (D) electrode and the voltage is measured between the $V$+ and $V$- electrodes. (h)-(i) The $V_{xx}^{1\omega}$ and $V_{xx}^{2\omega}$ as a function of a.c. current amplitude $I_{ac}$ measured under **B** = 9 T and $T$ = 10, 20, 30, 40, and 80 K, respectively.

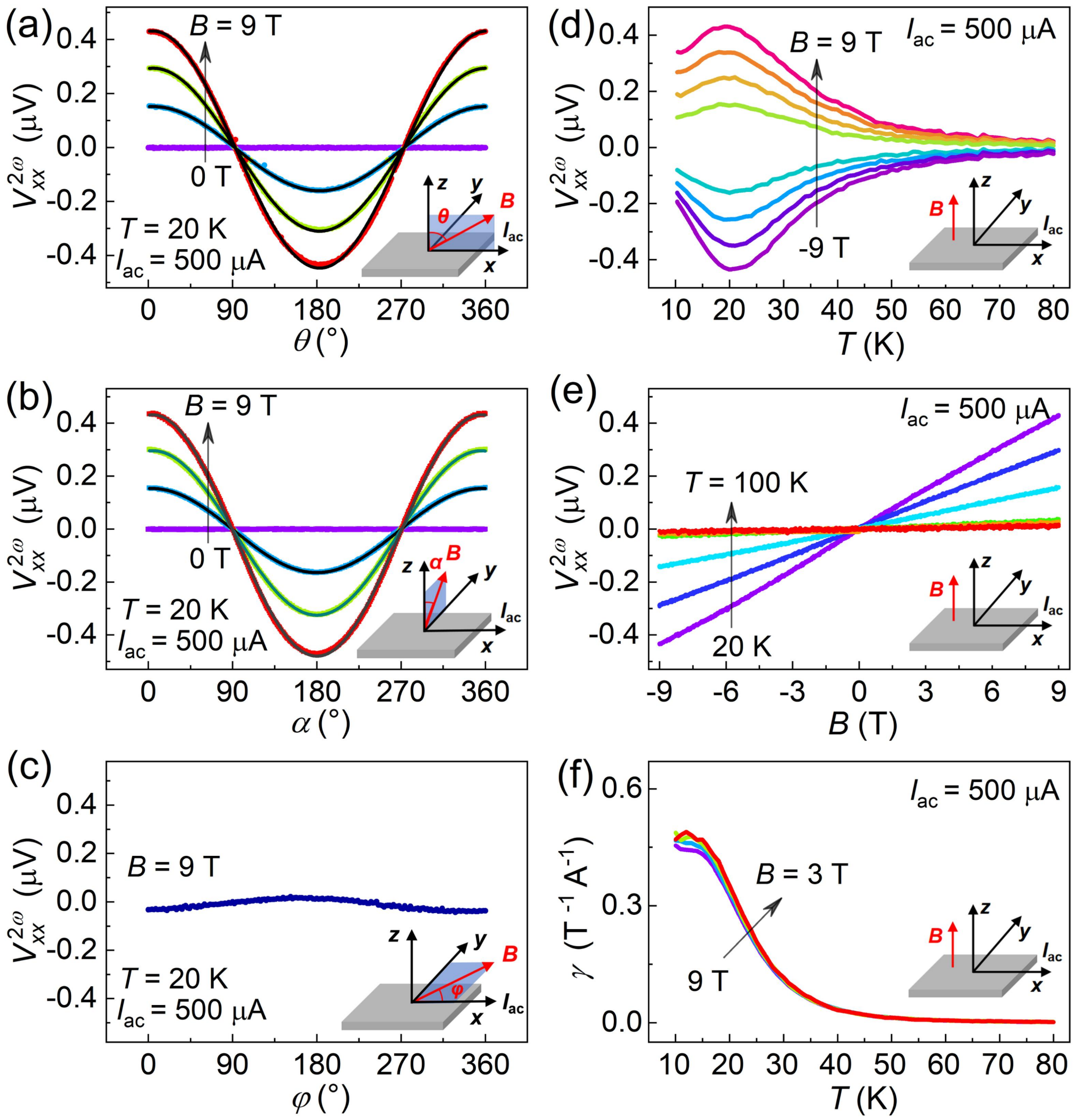

FIG. 3. Angular, temperature, and field dependence of the electrical magnetochiral anisotropy (eMChA). (a)-(c) Magnetic-field-orientation dependent $V_{xx}^{2\omega}$ at $T = 20$ K, $I_{ac} = 500$ $\mu$A and in **B** = 0, 3, 6, and 9 T (except for (c) where field **B** = 9 T) as the magnetic field was rotated in the *xz*, *yz*, and *xy* planes. The plane of rotation and the definition of the angles ($\theta$, $\alpha$ and $\varphi$) are shown in the insets of each panel. (d) Temperature dependent $V_{xx}^{2\omega}$ measured at $I_{ac} = 500$ $\mu$A in vertical **B** = 9, 7, 5, 3, -3, -5, -7 and -9 T. The inset shows the electrode geometry. (e) Magnetic-field dependent $V_{xx}^{2\omega}$ measured at $I_{ac}$ = 500 $\mu$A and $T$ = 20, 30, 40, 70, 80 and 100 K. The inset shows the electrode geometry. (f) Temperature dependent magnetochiral anisotropy coefficient γ at $I_{ac} = 500$ $\mu$A and **B** =3, 5, 7 and 9 T. The inset shows the electrode geometry. Here the magnetochiral anisotropy coefficient $\gamma = \frac{2R^{2\omega}}{R^{1\omega}|B||I|}$, extracted from Supplementary Figs. S4(a) and S4(b).

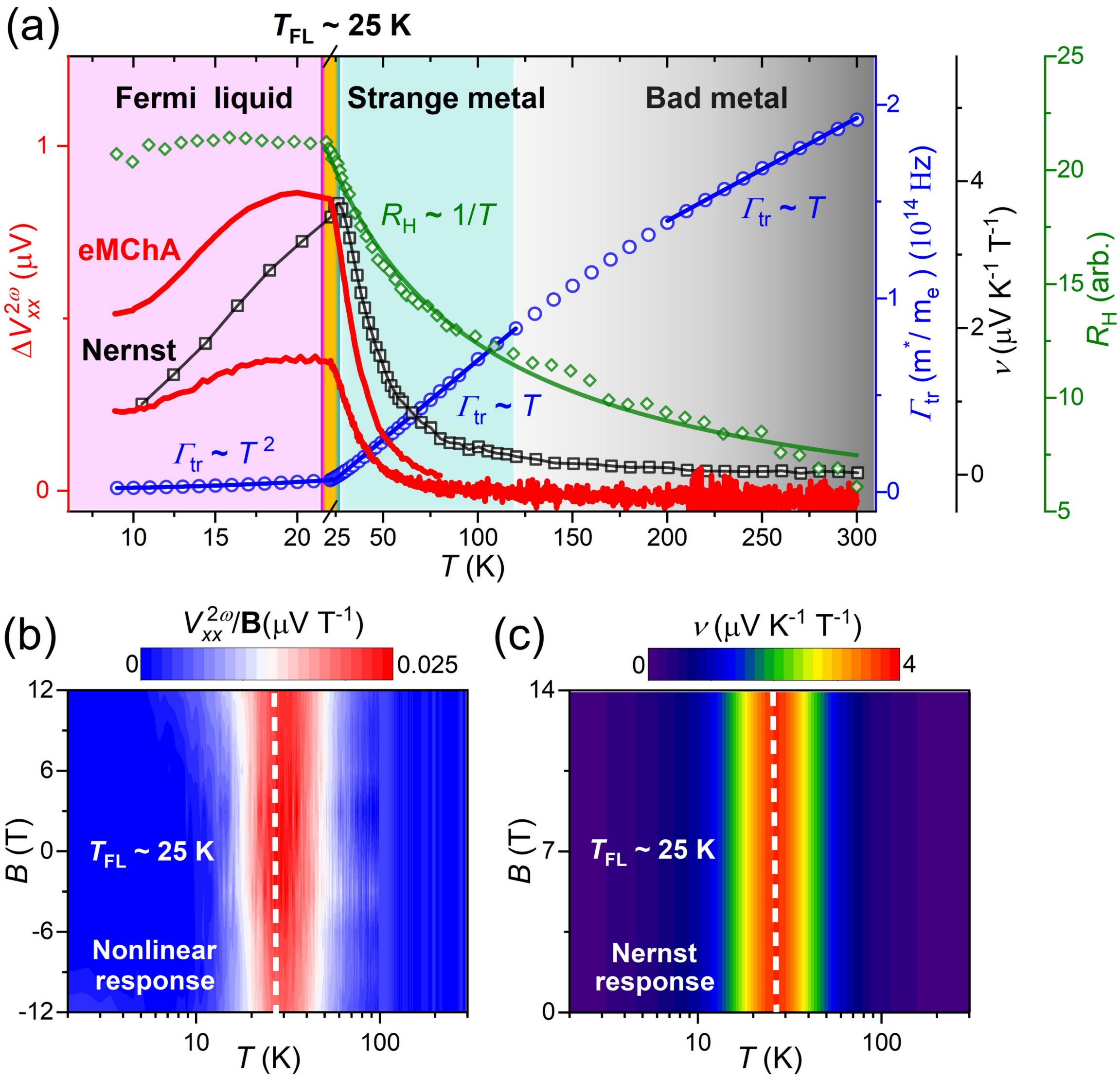

FIG. 4. Hidden symmetry breaking and intertwined physics in the normal state of 2$M$-$WS_2$ around $T$ ~ 25 K. (a) Two solid red lines represent temperature dependent second-harmonic voltage $\Delta V_{xx}^{2\omega}$=$\Delta V_{xx}^{2\omega}$(9 T)-$\Delta V_{xx}^{2\omega}$(-9 T) measured under $I_{ac}$ = 200 and 500 $\mu$A, respectively (extracted from Supplementary Figs. S4(b) and S4(c)). The black squares in the phase diagram represent the temperature dependent Nernst coefficient ν, extracted from Ref. [9]. The black line across the black squares is a guide to the eye. The blue circles represent the temperature dependent scattering rate extracted from resistivity (extracted from Supplementary Figs. S11 and S12), denoted as $\Gamma_{tr}(m^*/m_e)$. The factor of $m^*$ absorbs the correction of effective mass $m^*$, which is not determined in our experiments, where $m_e$ is the bare electron mass (see Supplementary Figs. S11 and S12 for details). The green diamonds represent temperature dependent Hall coefficient $R_H$ of 2$M$-$WS_2$ (extracted from Supplementary Fig. S11). The green solid line is the fitting of the temperature dependent $R_H$ to ~1/T form 25 K to 300 K, with very good fitting from 25 K to 120 K (strange metal region). Blue lines across the blue circles are quadratic and linear fittings at various sections of temperature ranges. The pink, light green, and gray shaded region in the phase diagram represent the Fermi liquid, strange metal and bad metal regions, respectively. The vertical yellow ribbon highlights the characteristic quasiparticle coherent temperature $T_{FL}$ ~ 25 K, at which the crossover from Fermi liquid to strange metal, the anomalous enhancement of large Nernst effect and the prominent magnetochiral anisotropy (eMChA) simultaneously take place in 2*M*-$WS_2$. To better highlight the two distinct regions as well as the FL-SM transition in Fig. 4(a), the break in the horizontal axis at $T$ = 22 K is introduced, which does not change the intrinsic physical information conveyed by the different data curves. (b) A contour plot of the magnetic field and temperature dependent normalized second-harmonic voltage $V_{xx}^{2\omega}/\mathbf{B}$ in a 2*M*-$WS_2$ flake, extracted from Supplementary Fig. S7(b). The white dashed line highlights the $T_{FL}$ ~ 25 K. (c) A contour plot of the magnetic field and temperature dependent Nernst coefficient ν in bulk 2*M*-$WS_2$, extracted and redrawn from Ref. [9]. The white dashed line highlights the $T_{FL}$ ~ 25 K.